# Absolute frequency measurements with a robust, transportable $^{40}$Ca$^+$ optical clock[*]

Huaqing Zhang[1,2], Yao Huang[1,2], Baolin Zhang[1,2], Yanmei Hao[1,2,3], Mengyan Zeng[1,2], Qunfeng Chen[1,2], Yuzhuo Wang[4], Shiying Cao[4]，Yige Lin[4]，Zhanjun Fang[4†], Hua Guan[1,2,5†], and Kelin Gao[1,2†]

[1]*State Key Laboratory of Magnetic Resonance and Atomic and Molecular Physics, Innovation Academy for Precision Measurement Science and Technology, Chinese Academy of Sciences, Wuhan 430071, China*
[2]*Key Laboratory of Atomic Frequency Standards, Innovation Academy for Precision Measurement Science and Technology, Chinese Academy of Sciences, Wuhan 430071, China*
[3]*University of Chinese Academy of Sciences, Beijing 100049, China*
[4]*National Institute of Metrology, Beijing 100029, China*
[5]*Wuhan Institute of Quantum Technology, Wuhan 430206, China*

We constructed a transportable $^{40}$Ca$^+$ optical clock (with an estimated minimum systematic shift uncertainty of $1.3 \times 10^{-17}$ and a stability of $5 \times 10^{-15}/\sqrt{\tau}$ ) that can operate outside the laboratory. We transported it from the Innovation Academy for Precision Measurement Science and Technology, Chinese Academy of Sciences, Wuhan to the National Institute of Metrology, Beijing. The absolute frequency of the 729 nm clock transition was measured for up to 35 days by tracing its frequency to the second of International System of Units. Some improvements were implemented in the measurement process, such as the increased effective up-time of 91.3 % of the $^{40}$Ca$^+$ optical clock over a 35-day-period, the reduced statistical uncertainty of the comparison between the optical clock and hydrogen maser, and the use of longer measurement times to reduce the uncertainty of the frequency traceability link. The absolute frequency measurement of the $^{40}$Ca$^+$ optical clock yielded a value of 411042129776400.26 (13) Hz with an uncertainty of $3.2 \times 10^{-16}$, which is reduced by a factor of 1.7 compared with our previous results. As a result of the increase in the operating rate of the optical clock, the accuracy of 35 days of absolute frequency measurement can be comparable to the best results of different institutions in the world based on different optical frequency measurements.

## 1. Introduction

The second of International System of Units (SI) is referenced to the unperturbed ground-state hyperfine transition frequency (9192631770 Hz) of the $^{133}$Cs atom and is realised by a set of primary and secondary frequency standards (PSFS) that are located in different national metrological laboratories around the world. Taking advantage of the petahertz-level ($10^{15}$ Hz) optical clock transitions, some state-of-the-art, optical atomic clocks surpassed the most advanced $^{133}$Cs fountain microwave clocks [1–4] by two orders of magnitude both in terms of uncertainty and stability [5–10]. Optical atomic clocks will be very likely used to redefine the SI second in the future. Meanwhile, optical clocks (OCs) are potential tools in applied physics and fundamental science and can be used extensively, including the tests of fundamental physics constants [11–13] and general relativity [6,14], gravitational wave and dark matter detection [15,16], and geodesy and metrology [17–20]. In addition, the use of OCs to steer the hydrogen maser (HM) to generate higher precision timescales than the traditional $^{133}$Cs fountain microwave clock constitutes another useful application [21,22].

---

[*] **Corresponding authors:** zfang@nim.ac.cn (Zhanjun Fang), guanhua@apm.ac.cn (Hua Guan), klgao@apm.ac.cn (Kelin Gao)

To-this-date, 10 optical frequency standards have been recommended as secondary representations of the second by the Comité International des Poids et Mesures (CIPM). The estimated systematic uncertainties of these OCs are less than $1 \times 10^{-16}$ and the uncertainties of the absolute frequencies are less than $2 \times 10^{-15}$. To redefine the SI second referenced to OCs in the future, the CIPM provided advice and guidelines for OCs [23]. One of them involves the frequency comparison measurements between OCs and $^{133}$Cs fountain clocks, which are limited by the uncertainty of the latter (e.g. $\Delta v/v < 3 \times 10^{-16}$). To engage in the definition of SI seconds in the future, solutions that were used to improve the accuracy of absolute frequency measurements of the $^{40}$Ca$^+$ OC ought to be revisited.

In this study, some improvements have been implemented in the measurement process, including the increase of the effective up-time of the optical clock to 91.3 %, the use of ultra-low phase noise frequency synthesiser to reduce the statistical uncertainty of the frequency comparison between the OC and HM, and the increase of the total absolute frequency measurement period up to 35 days to reduce the uncertainty of the frequency traceability link. The absolute frequency measurement outcome for the $^{40}$Ca$^+$ OC is 411042129776400.26 (13) Hz, and the uncertainty is $3.2 \times 10^{-16}$, which is reduced by a factor of 1.7 compared with recent results [20,24].

2. **Experimental setup and the improvement of the transportable $^{40}$Ca$^+$ OC**

The absolute frequency measurement of the OC can be compared directly with the local caesium primary standard or the frequency link to the International Atomic Time (TAI) via satellites. As is shown in figure 1, the transportable $^{40}$Ca$^+$ OC is located outside building #22 at the National Institute of Metrology (NIM), and the clock laser is sent to the optical frequency comb (OFC) laboratory by a 100 m single-mode polarization maintaining (PM) fiber using active fiber noise cancellation (FNC). The $^{133}$Cs fountain clock and HM are located at building #20 (NIM), and the 10 MHz frequency signal of the HM is sent to the OFC lab by an optical-microwave frequency transmission link. The absolute frequency of the transportable $^{40}$Ca$^+$ OC is measured by using a frequency link to the TAI to provide traceability to the SI second [25]. First, the frequency of the $^{40}$Ca$^+$ optical clock is compared with UTC (NIM) locally based on the use of an OFC and a flywheel HM to obtain the optical clock frequency referenced to UTC (NIM). The frequency deviation between UTC and UTC (NIM) can then be obtained from Circular T (published by the International Bureau of Weights and Measures (BIPM) every month), and the frequency of the $^{40}Ca^+$ OC can be associated with UTC or the TAI (there is a fixed time difference between UTC and the TAI). The frequency accuracy of the TAI is then evaluated by comparing the TAI scale unit based on calibrations of the SI second produced by primary and secondary frequency standards(PSFS).

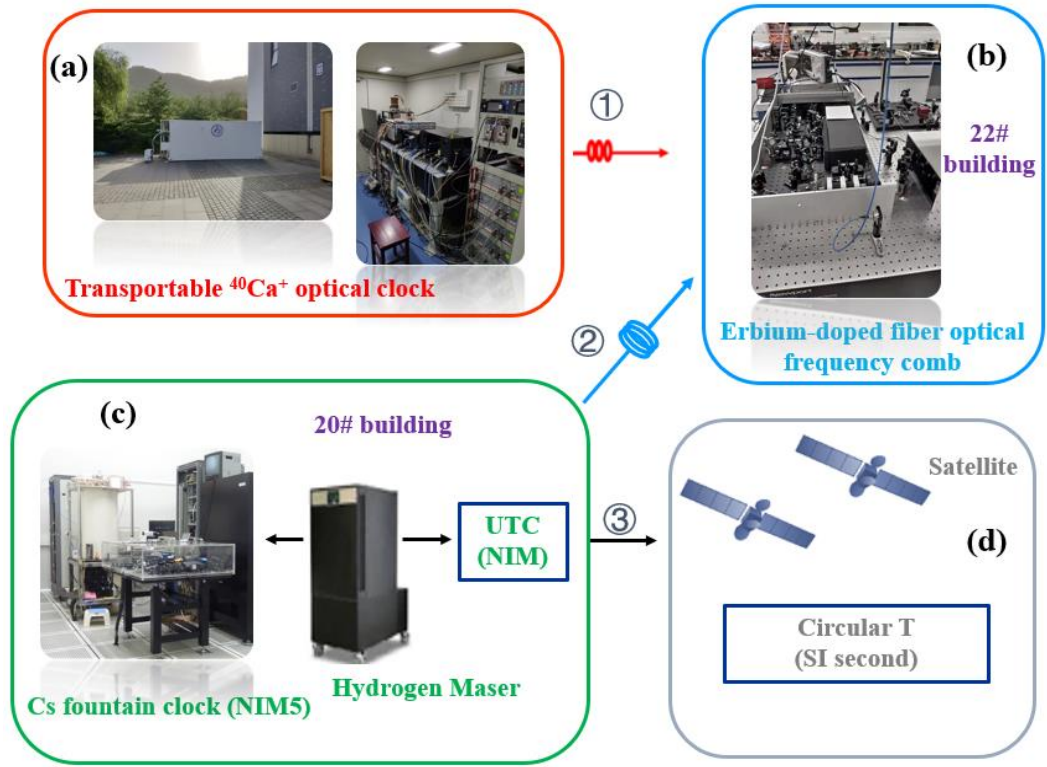

Figure 1. Schematic of absolute frequency measurement of the $^{40}Ca^+$ optical clock (OC). (a) Exterior and interior structures of the transportable $^{40}Ca^+$ OC. (b) The Erbium-doped fiber optical frequency comb (OFC) is used to compare the frequencies between the OC and the hydrogen maser (HM). (c) Cs fountain clock (NIM5) and HM ensembles are used to generate the UTC (NIM). (d) Circular T is published by the International Bureau of Weights and Measures every month and enables the use of the frequency traceability link to the realization of SI second. ① 100 m polarization maintaining (PM) fiber for 729 nm clock laser, ② optical-microwave transmission link, ③ frequency link to the International Atomic Time (TAI) using satellites.

The transportable $^{40}Ca^+$ OC (figure 1, upper left) was installed in an air-conditioned cabin [20,26]. It was first moved from a laboratory in Wuhan and the geopotential was measured based on frequency comparisons with the stationary $^{40}Ca^+$ OC [20]. The optical system, including three laser boxes (laser distribution and frequency stabilization) and the vacuum chamber of the ion trap, was installed on a 2.4 × 0.9 m$^2$ optical platform; the electronics system was installed in two 19-in equipment cabinets.

To ensure that the $^{40}Ca^+$ OC operates reliably outdoors, it is most important to provide it with a stable temperature and low-vibration noise environment. The OC was housed inside the air-conditioned cabin. To make the temperature inside the compartment more stable, especially the environment around the vacuum system that traps the ion (which affects the frequency shift of the blackbody radiation (BBR) of the OC), we added proportional control to the air conditioner. A thermistor was installed near the ion trapping vacuum system. Furthermore, according to the temperature of the thermistor, the air conditioner was controlled to be turned on and off. In addition, the cooling or heating mode of the air conditioner can also be used to control the temperature of the room. This temperature control ensures that the transportable $^{40}Ca^+$ clock can operate in outdoor temperatures in the range of -20–40 °C.

Another aspect that requires attention is the vibration environment. The four outer corners of the cabin were equipped with levelling mechanisms to maintain the cabin levelling without being affected by the ground outside the cabin. In addition, the optical platform inside the cabin was also equipped with a levelling mechanical structure to ensure the level of the optical system of the $^{40}Ca^+$ OC. Regarding mechanical shock absorption, the optical system of the OC was installed on an optical platform and steel wire springs were used for shock absorption under the optical platform. The circuit control system was spatially housed in two equipment racks, which were installed and cushioned with rubber at the posterior and bottom parts. We then separated the outer compressor of the air conditioner from the cabin (we placed the compressor of the air conditioner on the ground after we arrived at the experimental site). The final vibration level of the cabin environment was close to the vibration level of the laboratory.

After we moved to the NIM, the system uncertainty and stability were re-evaluated. Table 1 lists the uncertainty budget of the transportable $^{40}Ca^+$ OC. The main system uncertainty is attributed to the temperature-induced black body radiation (BBR) frequency shifts.

$$\Delta v_{BBR,i} = -\frac{1}{2h}(831.943 V/m)^2 \left(\frac{T_{BBR}(K)}{300K}\right)^4 \alpha_{0,i}(1+\eta_i), i = e, g \quad (1)$$

As shown in formula (1) [27], there are three items that affect the BBR shift: temperature $T_{BBR}(K)$, static scalar polarizability $\alpha_{0,i}$ and the dynamic correction factor $\eta_i$ ($\eta_i$ is negligible when calculating the BBR shifts). In a previous study, we measured the magic trapping radiofrequency (RF) ($\Omega = 2\pi \times 24.801$ (2) MHz) for the $^{40}Ca^+$ optical clock and accurately obtained the differential static scalar polarizability $\Delta\alpha_0$ of the clock transition [27]. The uncertainty of static polarizability affects the calculated frequency shift of BBR and is at the fractional $10^{-19}$ level [27]. Regarding the evaluation of BBR shifts induced by temperature, we used the peak-to-peak temperature value of the vacuum chamber change over a month as the uncertainty of the vacuum chamber temperature. At the same time, an infrared thermal imaging camera was used to detect the temperature after the RF trapping frequency was turned on and off, and the temperature change caused by RF heating was evaluated [28]. The actual temperature of the vacuum chamber and ion trap structure were 292.83 (0.99) K and 293.08 (1.24) K, respectively. Finally, according to their contributions to the solid angle of the ion, the temperature experienced by ions was calculated to be 292.93 (1.09) K. The materials of the vacuum cavity and ion trap were polished metal materials and their surface emissivities were less than one. We treated the entire system as a black body (surface emissivity was set to one) to magnify the uncertainty of temperature evaluation for a conservative evaluation, and the solid angle of thermal radiation received by ion was calculated using the geometric solid angle. The BBR shift induced by the fluctuation of temperature was estimated to be $1.2 \times 10^{-17}$.

Table 1. System uncertainty budget of the transportable $^{40}Ca^+$ optical clock

| Contributors | Frequency shifts (mHz) | Uncertainty (mHz) | Relative uncertainty (10$^{-17}$) |
|---|---|---|---|
| Blackbody radiation (BBR) Temperature | 345 | 5.1 | 1.2 |
| BBR coefficients including dynamic correction | -0.4 | <0.1 | <0.01 |
| Excess micromotion | 0 | <0.1 | <0.01 |
| Second-order Doppler - secular motion | -6.0 | 0.7 | 0.17 |
| AC Stark | 1 | 2 | 0.5 |
| Residual quadrupole | 0 | <0.1 | 0.01 |
| Residual 1$^{st}$ order Zeeman | 0 | 0.2 | <0.01 |
| Second order Zeeman | <0.1 | <0.1 | <0.01 |
| AOM chirping | 0 | <0.1 | <0.01 |
| Line pulling | 0 | <0.1 | <0.01 |
| Collision | 0 | <0.1 | <0.01 |
| First-order Doppler | 0 | <0.1 | <0.01 |
| Servo | 2.5 | 0.25 | <0.01 |
| Total | 342.1 | 5.5 | 1.3 |

The micromotion frequency shift (including excess micromotion) and the AC stark frequency shift (scalar polarization term) caused by the RF electromagnetic field cancel each other by using the magic RF frequency to trap the $^{40}Ca^+$ ion, and the uncertainty is at the $10^{-19}$ level [27]. The electrical quadrupole frequency shift and the first-order Zeeman shift are also cancelled by averaging the frequencies of several pairs of Zeeman components [29]. In addition, we used a fast Field Programmable Gate Array (FPGA)control system to lock the OC. Some system frequency shifts, such as servo errors, residual first-order Zeeman frequency shifts, and the ion's temperature induced second order Doppler shifts were optimised [30,31]. The total system uncertainty of the transportable $^{40}Ca^+$ OC was $1.3 \times 10^{-17}$.

The clock laser was frequency locked to a transportable 30 cm ultra-low expansion glass (ULE) cavity and the stability was less than $2 \times 10^{-15}$ @ 1–10 s (figure 2a, green circles). Figure 2a (red-dashed line) shows that the stability of the $^{40}Ca^+$ OC is $5 \times 10^{-15}/\sqrt{\tau}$ based on the use of a time-interleaved self-comparison method (by locking two pairs of different Zeeman split levels, a Fourier-limited spectral width of 10 Hz was obtained by using an interrogation pulse width of 80 ms). The stability of the transportable $^{40}Ca^+$ OC was verified based on frequency comparisons with another $^{87}Sr$ optical lattice clock located in NIM with a stability of $7 \times 10^{-15}/\sqrt{\tau}$ (figure 2a (pink circle and dashed-line data)). By using the automatic locking algorithm [30], the clock can auto-lock by assessing different Rabi spectrum line-widths using the clock interrogation times of 2 ms, 10 ms, and 40 ms in instances in which the clock underdoes sudden magnetic field mutations or vibrational shock. The total effective up-time of the

$^{40}Ca^+$ OC was increased to 91.3 % during the 35-day period (figure 2b).

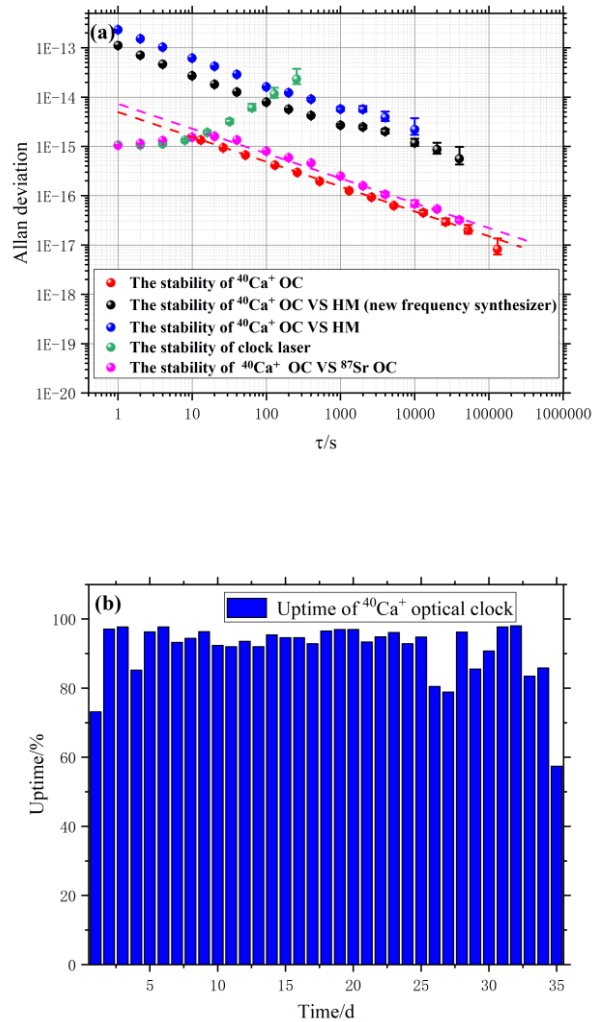

Figure 2. (a) Stability of the transportable $^{40}Ca^+$ OC using the time interleaved self-comparison (red circles). Stability of frequency comparisons between the $^{40}Ca^+$ OC and the HM (blue and black circles) using a frequency counter (Agilent, 53230A). Stability was improved by using a new, low-phase noise frequency synthesiser (black circles). The green circle data show the stability of the 729 nm clock laser. The pink circle data show the stability of frequency comparison between the transportable $^{40}Ca^+$ OC and the stationary $^{87}Sr$ optical lattice clock in the NIM's lab. (b) Daily effective uptime of the transportable $^{40}Ca^+$ OC.

## 3. Uncertainty evaluation for the absolute frequency measurement of transportable $^{40}Ca^+$ OC

The overall uncertainty for absolute frequency measurements includes the systematic uncertainty of the OC (described in section 2.2), statistical uncertainty of the frequency comparison between the OC and HM, uncertainty due to the frequency traceability link from the HM frequency to the SI second, systematic uncertainty of the OFC, and the gravitational redshift uncertainty. In our previous study [20], the frequency traceability link uncertainty attributed to the HM frequency tracing to the SI second was the dominant contributor to the overall uncertainty of the absolute frequency measurement and was limited by the measurement time of OC (an uptime rate >75 % over a 20-day period). In this study, we reduced the uncertainty from $5 \times 10^{-16}$ to $2.5 \times 10^{-16}$ (see section 3.2) by increasing the continuous running time and operating rate of the optical clock to 35 days and 91.3 %, respectively. The statistical uncertainty of the frequency comparison between the OC and HM was also reduced from $2.5 \times 10^{-16}$ to $0.9 \times 10^{-16}$ owing to the increased running time and operation rate of the OC and the frequency synthesiser with low-phase noise used as the reference of the repetition frequency $f_r$ of the OFC (see section 3.1).

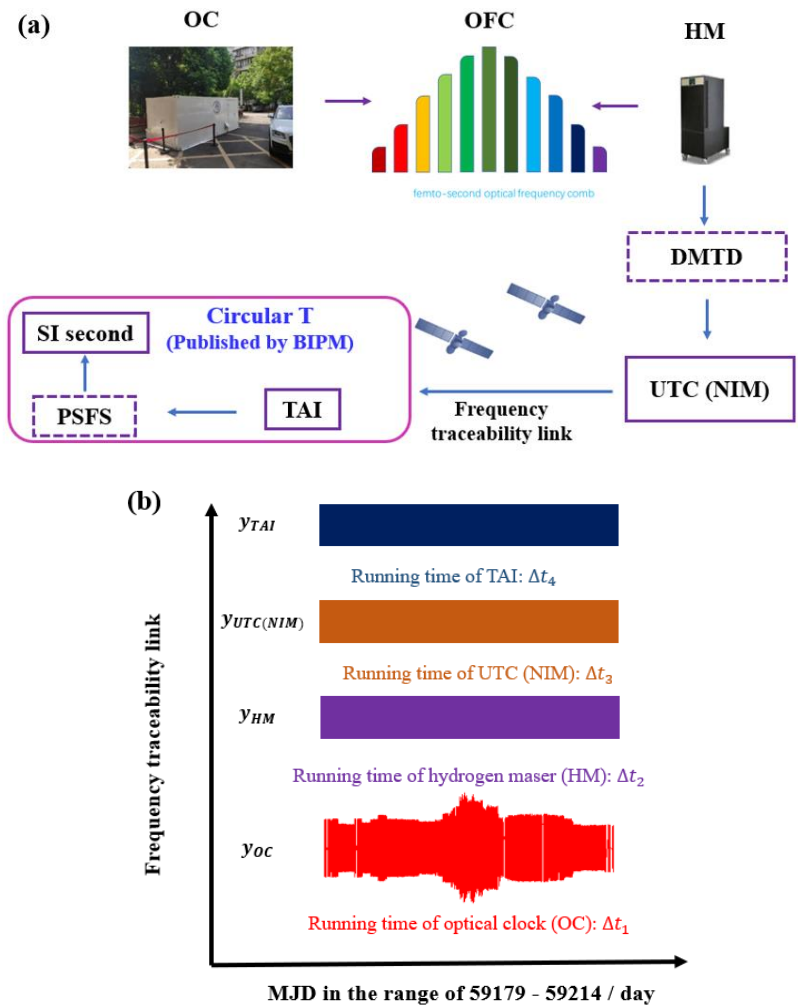

Figure 3. (a) Absolute frequency traceability to SI seconds for $^{40}Ca^+$ OCs. The dual-mixer time difference measurement system (DMTD) was used to perform frequency comparisons between the HM and UTC (NIM). (b) Running time of responses of all measurement periods when using the frequency traceability link. The shaded area (red colour) corresponds to the real-time data from the frequency comparison of OC and HM and shows the increased up-time of the OC. The running time

of the HM, UTC (NIM), and the International Atomic Time (TAI) are 100 % during the modified Julian dates (MJD) in the range of 59179–59214, and the purple, brown, and blue-shaded areas symbolise the running conditions of HM, UTC (NIM), and TAI, respectively.

The absolute frequency traceability to SI second for the $^{40}Ca^+$ OC (figure 3) was divided into four parts: the frequency comparisons between OC and HM, HM and local UTC (NIM), UTC (NIM) and TAI, and the frequency comparison between TAI and SI second. As shown in formula (2), the frequency ratio of OC vs. SI second can be written as,

$$\frac{y_{OC}}{y_{SI \cdot s}} = \frac{y_{OC}}{y_{HM}} \times \frac{y_{HM}}{y_{UTC(NIM)}} \times \frac{y_{UTC(NIM)}}{y_{TAI}} \times \frac{y_{TAI}}{y_{SI \cdot s}} \qquad (2)$$

However, the OC and OFC do not operate continuously in actual conditions; thus, the comparison between the OC and HM involves a dead time (outside the operating time period of the OC). We need to deduce the frequency offset caused by the dead time from the comparison data of OC and HM. At the same time, the time intervals from UTC (NIM) to TAI must also be aligned. If not, the frequency offset due to data misalignment must also be considered. Considering the dead time, formula (2) also can be written as,

$$\frac{y_{OC}}{y_{SI \cdot s}} = \frac{y_{OC}(\Delta t_1)}{y_{HM}(\Delta t_1)} \times \frac{y_{HM}(\Delta t_1)}{y_{HM}(\Delta t_2)} \times \frac{y_{HM}(\Delta t_2)}{y_{UTC(NIM)}(\Delta t_2)} \times \frac{y_{UTC(NIM)}(\Delta t_2)}{y_{UTC(NIM)}(\Delta t_3)} \times \frac{y_{UTC(NIM)}(\Delta t_3)}{y_{TAI}(\Delta t_3)} \times \frac{y_{TAI}(\Delta t_3)}{y_{TAI}(\Delta t_4)} \times \frac{y_{TAI}(\Delta t_4)}{y_{SI \cdot s}} \qquad (3)$$

where $\Delta t_i$ ($i = 1, 2, 3, 4$) denotes the running time of each measurement. The uptime of OFC is 85 % during this measurement period (Circular T 396, Date: 2020.11.26 8:00–2020.12.31 8:00; MJD: 59179–59214), while the HM, UTC (NIM), and TAI were operating at the uptime of 100 % ($\Delta t_2 = \Delta t_3 = \Delta t_4$).

*3.1. Frequency comparison between the $^{40}Ca^+$ OC and HM*

The OFC can establish the association between microwave and optical frequencies. Using the OFC, we can measure directly the ratio of an OC to a HM. Figure 4 shows the $^{40}Ca^+$ OC frequency measurement referenced to HM. The data also show the high-running rate of 85 % when considering the uptime of OFC. We counted the measured data in days and derived a mean value $x_i$ and a standard deviation $\sigma_i$ from the daily data. The final measurement result (MJD: 59179–59214) was calculated according to the weighted average variance, and its absolute value was 4110421297764 24.272 (36) Hz. The statistical uncertainty of $\frac{y_{OC}(\Delta t_1)}{y_{HM}(\Delta t_1)}$ was optimised from $2.5 \times 10^{-16}$ to $0.9 \times 10^{-16}$ owing to the increased continuous operation rate for periods up to 35 days. By using the weighted linear fit of the data in figure 4b, we derived the linear frequency drift of the HM (VCH-1003M Option L, No. 4850) to be equal to $5.6 \times 10^{-17}$/day. The drift rate was consistent with that of the HM on the BIPM website.

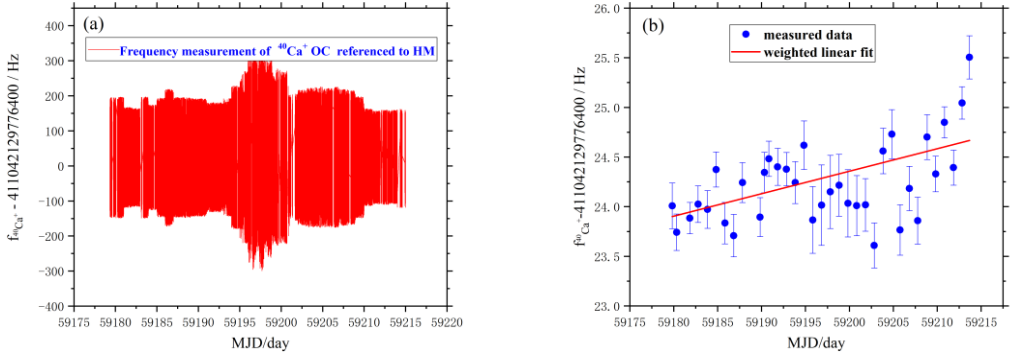

Figure 4. $^{40}Ca^+$ optical clock frequency measurement referenced to HM. (a) The measured data show the increased running time of 85 % when the up-time of the OFC was considered. (b) The measured data were averaged on a daily basis and the red solid line shows that the linear frequency drift of the HM is $5.6 \times 10^{-17}$/day.

*3.2. Uncertainty evaluation of HM frequency traceability link to SI second*

As described in formula (3), the fractional frequency shift and uncertainty of each part are listed in Table 2 when the frequency of the OC traces back to SI seconds. Detailed evaluation methods are discussed in the following sections.

Taking advantage of an OFC, we used a highly stable, continuously running HM (figure 2a, black circle data) as the flywheel oscillator to correlate the frequency of the OC with UTC (NIM). As the optical clock has a dead time, two points must be considered when the OC frequency is compared with the HM: first, the frequency offset caused by the time alignment between OC and the HM; second, the frequency offset caused by the death time of the OC. We first estimated the linear frequency drift rate $\dot{f}_{HM}$ of the HM by comparing the OC and HM data (figure 4b). The midpoint up-time $\bar{t}_{OC}$ of the OC and the midpoint up-time $\bar{t}_{HM}$ of the HM were then calculated. The frequency offset caused by the HM frequency drift during the frequency measurement of the OC was calculated based on the formula: $\dot{f}_{HM} \times (\bar{t}_{HM} - \bar{t}_{OC}) = 1(3) \times 10^{-17}$. For the frequency offset caused by the dead time of OC and OFC, we evaluated the noise model of HM by comparing the OC and HM data after we eliminated the linear frequency shift; we then evaluated the frequency offset caused by the dead time based on 100 simulation iterations according to the HM's noise model (the relevant method is in [32,33]). As the OC had been running every day for 35 days and the total operation rate reached 91.3 %, the uncertainty caused by the dead time of the OC and OFC was reduced to $1 \times 10^{-16}$. The total frequency calibration of $\frac{y_{HM}(\Delta t_1)}{y_{HM}(\Delta t_2)}$ was estimated to be $0.1\,(1) \times 10^{-16}$.

We used the dual-mixer time difference (DMTD) measurement system to perform frequency comparisons and acquire the HM and UTC (NIM) time difference. The frequency calibration $\frac{y_{HM}(\Delta t_2)}{y_{UTC(NIM)}(\Delta t_2)} = -4.5238 \times 10^{-14}$ and the statistical uncertainty were estimated to be $5 \times 10^{-17}$ after 35 days according to a zero measurement using a common signal introduced to DMTD [34]. The time difference between UTC (NIM) and TAI can be found in Circular T 396 (published by the BIPM). The frequency calibration $\frac{y_{\text{UTC(NIM)}}(\Delta t_3)}{y_{\text{TAI}}(\Delta t_3)} = -5.29(1.7) \times 10^{-16}$. According to Circular T 396, we can acquire the estimated frequency shift of $d = 1.9 \times 10^{-16}$ by the BIPM based on all PSFS measurements identified to be used for TAI steering over the period MJD 59179-59214, and corresponding uncertainties $u = 1.4 \times 10^{-16}$. The frequency calibration yielded $\frac{y_{\text{TAI}}(\Delta t_4)}{y_{\text{SI s}}(\Delta t_4)} = 1.9(1.4) \times 10^{-16}$.

As described in Table 2, the total calibration values of frequency traceability link was estimated to be $-455.67(2.5) \times 10^{-16}$.

Table 2. Summary of calibration values of frequency traceability link

| NO. | Frequency traceability link | Shift | Uncertainty |
|---|---|---|---|
| 1 | $\frac{y_{\text{HM}}(\Delta t_1)}{y_{\text{HM}}(\Delta t_2)}$ | $1 \times 10^{-17}$ | $1 \times 10^{-16}$ |
| 2 | $\frac{y_{\text{HM}}(\Delta t_2)}{y_{\text{UTC}(NIM)}(\Delta t_2)}$ | $-4.5238 \times 10^{-14}$ | $5 \times 10^{-17}$ |
| 3 | $\frac{y_{\text{UTC}(NIM)}(\Delta t_2)}{y_{\text{UTC}(NIM)}(\Delta t_3)}$ | 0 | 0 |
| 4 | $\frac{y_{\text{UTC}(NIM)}(\Delta t_3)}{y_{\text{TAI}}(\Delta t_3)}$ | $-5.29 \times 10^{-16}$ | $1.7 \times 10^{-16}$ |
| 5 | $\frac{y_{\text{TAI}}(\Delta t_3)}{y_{\text{TAI}}(\Delta t_4)}$ | 0 | 0 |
| 6 | $\frac{y_{\text{TAI}}(\Delta t_4)}{y_{\text{SI s}}(\Delta t_4)}$ | $1.9 \times 10^{-16}$ | $1.4 \times 10^{-16}$ |
| | Sum | $-4.5567 \times 10^{-14}$ | $2.5 \times 10^{-16}$ |

*3.3. System uncertainty of the OFC*

The optical comb used in the experiment was a homemade, erbium-doped, fiber femtosecond OFC [35] which was used to compare frequencies of the OC and HM. The measured frequency of the OC can be written as

$$f_{OC} = nf_r + f_o + f_b \tag{4}$$

where $n$ is an integer mode number which is about 2100000, $f_r$ is the repetition frequency, $f_0$ is carrier envelope offset frequency, and $f_b$ is the beat frequency at 729 nm of the $^{40}Ca^+$ clock transition.

The three parameters $f_r$, $f_0$, and $f_b$ were introduced in the calculation of the frequency of the $^{40}Ca^+$ OC. The frequency resolution and uncertainty of the microwave syntheses ($f_r$ and $f_0$) that were referenced to the HM need to be considered. The counter error caused by the frequency counter must also be considered.

To evaluate the counting error, we used the frequency counter (referenced to a 10 MHz HM signal) to measure the 10 MHz HM signal. The Allan deviation of the data was $\sigma_{counter} = 1.17 \times 10^{-12}/\sqrt{t}$; this outcome was limited by the resolution of the frequency counter (53230A, Agilent). The frequency beat signal $f_{b,Ca^+}$ and carrier envelope offset frequency $f_0$ were synchronously countered by two frequency counters and caused additive frequency offsets at optical frequencies. The counter error can be described by the following formula [36]

$$\sigma_{y,c} = \sigma_{counter} \times \frac{f_{input}}{f_{OC}} \tag{5}$$

The input frequency $f_0 = 20$ MHz and $f_{b,Ca^+} \approx 58$ MHz. Owing to the frequency of the OC $f_{OC} = 411$ THz, the counter error $\sigma_{y,c}$ was less than $1 \times 10^{-17}$.

Table 3. System uncertainty of the optical frequency comb (OFC)

| Items | Frequency shifts | Uncertainty |
|---|---|---|
| Counter ($f_b$ and $f_0$) | 0 | $< 1 \times 10^{-17}$. |
| Frequency synthesiser ($f_r$) | 0 | $1.6 \times 10^{-16}$ |
| Sum | 0 | $1.6 \times 10^{-16}$ |

As the mode number $n$ of the OFC at 729 nm is a large number ($n \approx 2\,100\,000$), a change of 1 µHz in the repetition frequency ($f_r \approx 195.4$ MHz) will result in a relative optical frequency change of $5 \times 10^{-15}$. It is of vital importance to control and measure $f_r$ accurately. The reference frequency of $f_r$ was generated by an ultra-low, phase noise frequency synthesiser (ROHDE & SCHWARZ, SMB 100B). According to technical notes, the normal resolution of synthesis was 0.163 µHz when the output frequency of the frequency synthesiser was set at 1 GHz; this corresponds to a relative uncertainty of $1.6 \times 10^{-16}$. Considering the errors of the frequency counter and the frequency synthesiser, the system uncertainty of the OFC was estimated to be $0\,(1.6) \times 10^{-16}$ (Table 3).

*3.4. Gravitational redshift*

The gravitational redshift $\delta_{GR} = g \times \Delta h/c^2$ must be considered when the clock is located at different altitudes $\Delta h$ above the sea level. The averaged local gravity acceleration $g = 9.801\ 236\ 21\ (2)\ \text{m/s}^2$, and $c$ is the speed of light. The altitude $\Delta h$ was measured in the dimensional metrology laboratory of the NIM and was traced to a levelling benchmark (which is referenced to the China 1985 national height datum) inside the Changping campus. The altitude $\Delta h$ was estimated to be 110.3 (2) m when the transportable $^{40}$Ca$^+$ OC was located in the Changping campus of the NIM. The uncertainty of the $\Delta h$ involves the leveling error and the vertical datum offset between the regional height datum and the global height datum [37]. Based on the method described in reference [38], the gravitational redshift was 4.944 (9) Hz.

*3.5. Measured results of the absolute frequency of the $^{40}$Ca$^+$ OC*

Table 4. Summary of frequency correction for the absolute frequency measurement of the $^{40}$Ca$^+$ optical clock (OC). The frequency was subtracted by 411042129776400 (Hz)

| Frequency correction | Shift (Hz) | Uncertainty (Hz) | Relative uncertainty ($10^{-16}$) | Relative uncertainty of last measurement ($10^{-16}$) |
|---|---|---|---|---|
| Statistic (OC vs. hydrogen maser (HM)) | 24.272 | 0.036 | 0.9 | 2.5 |
| Traceability link | -18.730 | 0.103 | 2.5 | 5 |
| Gravitational redshift | -4.944 | 0.009 | 0.2 | 0.2 |
| OC | -0.3421 | 0.0055 | 0.13 | 1.3 |
| OFC | 0 | 0.066 | 1.6 | |
| Sum | 0.26 | 0.13 | 3.2 | 5.6 |

With the frequency correction described in sections 3.1, 3.2, 3.3, and 3.4, the absolute frequency of the $^{40}$Ca$^+$ OC was measured to be equal to 411042129776400.26 (13) Hz, with a frequency uncertainty equal to $3.2 \times 10^{-16}$. The contribution of each part is summarised in Table 4. The improvement of the measurement accuracy is due to the improvement of the measurement period (up to 35 days) and the up-time of OC (91.3 %) that reduced the uncertainty of traceability link. At the same time, the system uncertainty of the OFC was evaluated and verified.

4. **Comparison of the absolute frequency measurements of the $^{40}$Ca$^+$ OC**

Figure 5 shows the comparison of the absolute frequency measurements of the $^{40}$Ca$^+$ OC. The blue circle and error bar data denote the weighted averages calculated based on measurements from different research institutes (Innsbruck 2009 [39], NICT

2012[40], WIPM 2012 [41], and WIPM 2014/2015 [41]) and those published in BIPM 2017. The green circle and error bar data were measured by our group in January 2020 [20]. The purple circle and error bar data are the recommend values for the standard frequency of the $^{40}Ca^+$ OC published in BIPM 2021. The pink circle and error bar data were measured by another transportable $^{40}Ca^+$ OC housed in a laboratory [24]. The red circle and error bar data are our results (this study) that yielded an accuracy of $3.2 \times 10^{-16}$ which was comparable to the best results measured by different institutions based on different optical frequencies [36,42–46].

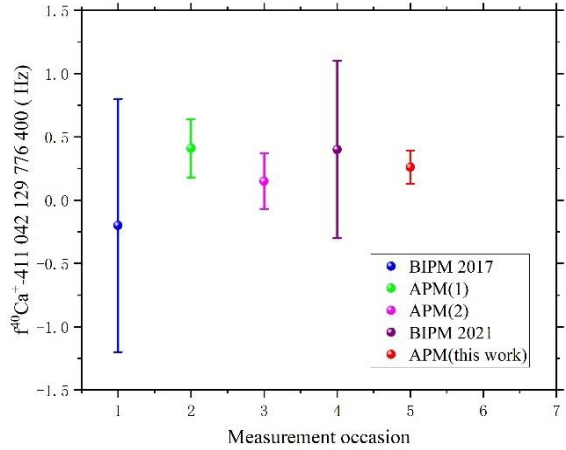

Figure 5. Comparison of the absolute frequency measurements of the $^{40}Ca^+$ OC

## 5. Conclusions and perspectives

In this study, we constructed a robust $^{40}Ca^+$ OC with a systematic uncertainty of $1.3 \times 10^{-17}$ and a stability of $5 \times 10^{-15}/\sqrt{\tau}$. The effective up-time rate was improved to 91.3 % over a 35-day-period after the use of the auto-locking scheme. The absolute frequency was measured in reference to remote PSFS. Furthermore, the stable operation of the OC for periods up to 35 days also greatly reduced the remote frequency link uncertainty of the OC frequency tracing to SI seconds.

The absolute frequency measurement value of the $^{40}Ca^+$ ion OC was 411 042129776400.26 (13) Hz and the uncertainty was $3.2 \times 10^{-16}$, which is the highest accuracy based on $^{40}Ca^+$ ion OC obtained to this date. The measured results were in good agreement with the previous measurements, and the reliability of the transportable $^{40}Ca^+$ ion OC was verified. Taking advantage of the high reliability and high-operation rate, it is possible to transport the OC to institutes that possesses HMs to frequency steer the HMs to produce high-precision time-scales. The frequency ratio measurements between the transportable $^{40}Ca^+$ ion OC and the stationary $^{87}Sr$ optical lattice clock locked in NIM constitute a part of our upcoming study. By improving the experimental scheme to reduce the BBR frequency shift of the transportable OC at the fractional $10^{-18}$ level, it is envisaged that the transportable clock will be also used in geodesic applications.

**Acknowledgments**

We thank Tianchu Li, Fang Fang for help and fruitful discussions and Huanyao Sun for his/her help with electronic circuits. This work was supported by the National Natural Science Foundation of China (Grants No. 12022414 and 12121004), the National Key R&D Programme of China (Grants No. 2018YFA0307500, 2022YFB3904001, 2021YFB3900701, 2021YFF0603802, and 2021YFF0600102), the Natural Science Foundation of Hubei Province (Grant No. 2022CFA013), the Chinese Academy of Sciences Youth Innovation Promotion Association (Grants No. Y201963 and 2018364), and the CAS Project for Young Scientists in Basic Research (Grant No. YSBR-055) . Huaqing Zhang and Yao Huang contributed equally to this work.